%% file: main.tex
\documentclass[conference]{IEEEtran}

\IEEEoverridecommandlockouts

\usepackage{makecell}
\usepackage{dingbat,utfsym}
\usepackage{cite}
\usepackage[hyphens]{url}
\usepackage{tikz}
\usepackage{braket}
\usepackage{subfig}
\usepackage{xcolor}
\usepackage{amsmath}
\usepackage{amssymb}
\usepackage{balance}
\usepackage{environ}
\usepackage{physics}
\usepackage{colortbl}
\usepackage{enumitem}
\usepackage{amsfonts}
\usepackage{booktabs}
\usepackage{graphicx}
\usepackage{textcomp}
\usepackage{algorithm}
\usepackage{algpseudocode}
\usepackage[normalem]{ulem}
\usepackage[most]{tcolorbox}
\usepackage{fontawesome5} 
\usepackage{pdfpages}
\usepackage{stmaryrd}
\usepackage{hyperref}

\usepackage{marginnote}

\def\BibTeX{{\rm B\kern-.05em{\sc i\kern-.025em b}\kern-.08em
    T\kern-.1667em\lower.7ex\hbox{E}\kern-.125emX}}

\newcommand{\sol}{Park-n-Ride}

\title{Hardware-Aware Compilation and Execution of Bivariate Bicycle Codes on Neutral-Atom Systems}

\author{\IEEEauthorblockN{Jason Ludmir}
\IEEEauthorblockA{\textit{Rice University}\\ Houston, TX, USA}
\and
\IEEEauthorblockN{Aditya Ranjan}
\IEEEauthorblockA{\textit{Northeastern University}\\ Boston, MA, USA}
\and
\IEEEauthorblockN{Nicholas S. DiBrita}
\IEEEauthorblockA{\textit{Rice University}\\ Houston, TX, USA}
\and
\IEEEauthorblockN{Jason Han}
\IEEEauthorblockA{\textit{Rice University}\\ Houston, TX, USA}
\and
\IEEEauthorblockN{Tirthak Patel}
\IEEEauthorblockA{\textit{Rice University}\\ Houston, TX, USA}
}

\begin{document}

\maketitle

\begin{abstract}

Quantum computers are noisy; without quantum error correction (QEC), deep programs fail as qubits lose information due to decoherence. Among QEC approaches, bivariate bicycle (BB) codes offer low overhead and constant-depth syndrome extraction, while neutral-atom arrays provide scalable, reconfigurable qubit layouts. However, executing BB-code primitives on neutral-atom systems requires a hardware-aware mapping that respects movement, zoning, and interaction constraints. We present \sol{}, a system for compiling and executing the BB code on neutral-atom processors. \sol{} introduces a module layout and movement model aligned with neutral-atom constraints, exposes a compact BB-native logical interface for compilation, and integrates scheduling mechanisms that enable efficient execution on zoned architectures. By co-designing BB-code abstractions with hardware execution, \sol{} provides a practical path from qLDPC primitives to resource-efficient, high-throughput execution on reconfigurable neutral-atom arrays.

\end{abstract}

\input{sections/introduction}
\input{sections/background}
\input{sections/motivation}
\input{sections/design}
\input{sections/methodology}
\input{sections/evaluation}
\input{sections/conclusion}

\balance

\bibliographystyle{IEEEtranS}
\bibliography{main}

\end{document}

%% file: sections/introduction.tex
\section{Introduction}
\label{sec:introduction}

Quantum error correction (QEC) is the key enabler for executing deep, long-running quantum programs in the presence of noise~\cite{holmes2020nisqboostingquantumcomputing,caliqec,lincodesign,Kim2024wob,yin2024qeccsynthlayoutsynthesizerquantum}. Among proposed codes, Quantum Low-Density Parity-Check (qLDPC) codes offer an attractive alternative to surface codes due to their constant-weight parity checks and constant degree under scaling~\cite{stein2024architecturesheterogeneousquantumerror, zhou2025lowoverheadtransversalfaulttolerance}. In particular, Bivariate Bicycle (BB) codes provide low overhead and high thresholds through a regular toroidal structure. On the other hand, neutral-atom systems have emerged as a leading platform for scalable quantum computation~\cite{cain2026shor}. They support large qubit counts, high-fidelity entangling gates via Rydberg interactions, and reconfigurable connectivity through qubit shuttling~\cite{saffman2010rmp,tan2022qubit,wang2024atomique}. These capabilities make them a natural substrate for qLDPC codes, whose non-local connectivity and regular structure are difficult to realize on fixed-layout architectures.

Despite this, executing BB codes efficiently on neutral-atom devices remains a systems challenge. BB codes require long-range interactions and toroidal wrap-around connectivity that do not map directly to zoned neutral-atom architectures. At the same time, neutral-atom operations are constrained by acousto-optic deflector (AOD)-based transport, non-crossing movement rules, and Rydberg blockade, which impose strict limits on how qubits can be moved and interacted with. Existing neutral-atom systems do not support BB-native operations, as these deviate from standard gate-based abstractions~\cite{liu2025coniqenablingconcatenatedquantum, salesrodriguez2025msd}. Enabling them to support BB codes is non-trivial due to the complexity of the codes and the system's constraints. As a result, there is currently no hardware-aware execution model for realizing BB codes on reconfigurable atom arrays.

We address this gap with \sol{}\footnote{The name reflects how BB modules are placed and moved in our setup. \sol{} is published in the Proceedings of the ACM/IEEE International Conference for High Performance Computing, Networking, Storage, and Analysis (SC), 2026.}, a hardware-software co-design framework for compiling and executing BB codes on neutral-atom systems. \sol{} introduces a BB-native execution model that captures movement, zoning, and interaction constraints~\cite{bluvstein2025architectural,bluvstein2021controlling,tan2025compilation}, enabling BB operations to be translated into valid, high-parallelism motion schedules. Neutral-atom systems combine large-scale qubit arrays, transport-enabled connectivity, and compatibility with qLDPC codes, while \sol{} supplies the missing hardware execution layer between BB logical compilation and physical neutral-atom constraints.

Our approach is driven by an HPC systems objective: maximize execution parallelism while respecting hardware constraints. \sol{} achieves this through three key components. (1) \textit{Spectral Placement:} a spectral seriation-based algorithm that places BB modules to minimize communication distance and movement overhead. (2) \textit{Parallel Shift Automorphisms:} a decomposition of BB shift operations into cyclic rolls with direction-aware scheduling that avoids AOD column crossings while maximizing concurrency. (3) \textit{Bridge-Based Interaction:} a scheduler for inter-module joint measurements that coordinates qubit movement within interaction zones while satisfying blockade and non-crossing constraints.

The contributions of this work are as follows:
\begin{itemize}
    \item A hardware-aware execution model for running BB-native operations on zoned neutral-atom systems with valid movement, interaction, and scheduling constraints.
    \item A parallel shift-automorphism scheduling strategy that avoids AOD column crossings and significantly reduces execution time relative to serialized approaches.
    \item A spectral seriation-based placement algorithm that reduces communication cost and runtime compared to arbitrary and greedy policies, and a bridge-based joint-measurement scheduler that guarantees conflict-free motion while exploiting available parallelism in the system.
    \item Evaluation using the $\llbracket 144,12,12\rrbracket$ gross code and the $\llbracket288,12,18\rrbracket$ two-gross code to demonstrate scalability. We also evaluate the effects of atom loss and provide end-to-end simulated timing results.
    \item Evaluation demonstrating up to $\sim$$40\%$ runtime reduction when scaling compute-column capacity (from 2 to 10), with consistently lower runtime than baseline strategies.
    \item \sol{}'s code and data are open-sourced at \textit{\url{https://github.com/positivetechnologylab/Park-n-Ride}}.
\end{itemize}

%% file: sections/background.tex
\section{Brief Relevant Background}
\label{sec:background}

\begin{figure}[t]
\centering
\includegraphics[width=0.98\columnwidth]{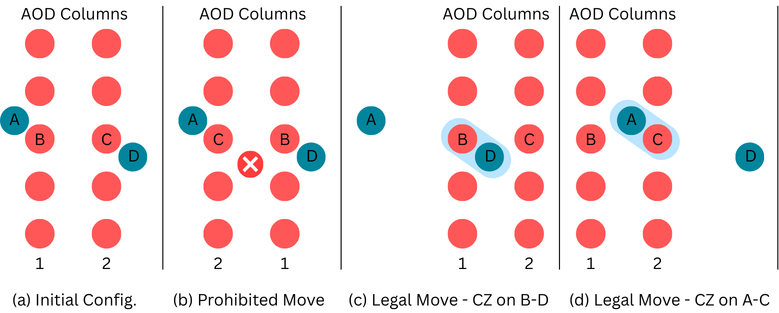}
\vspace{-1mm}
\caption{Acousto-optic deflector (AOD) non-crossing and stable column ordering. (a) Initial ordering $x_1<x_2<\dots$ fixed by AOD control. (b) Forbidden: executing CZ by moving column 2 past column 3 requires crossing. (c) Legal alternative: translate columns together to preserve ordering. (d) Even logically commuting CZs can conflict under non-crossing, forcing serialization.}
\vspace{-5mm}
\label{fig:aod}
\end{figure}

\subsection{Neutral Atom Hardware}
\label{sec:neutral atom-background}

\paragraph{Platform overview}
Neutral-atom processors trap atoms in optical tweezers and manipulate them using Raman or microwave drives. Two properties make them well-suited for large-scale execution: (1) reconfigurable geometry via programmable trap arrays and atom transport, and (2) strong, switchable interactions via Rydberg excitation, enabling high-fidelity two-qubit gates. Modern systems also support mid-circuit measurement and zoned operation over hundreds of qubits that enable error correction~\cite{saffman2010rmp,bluvstein2024logical,balewski2024engineering,sunami2025scalable}.

\paragraph{Trap technologies: SLM vs.\ AOD}
Atom arrays are realized using two complementary mechanisms. \emph{Spatial light modulators (SLMs)} generate static 2D layouts but do not support motion. \emph{AODs} enable continuous translation of atoms during execution. An \emph{AOD column} is a group of atoms translated together along a shared control channel. Let $x_c(t)$ denote the position of column $c$ at time $t$. The key constraint is \emph{non-crossing}: if $x_{c_1}(0) < x_{c_2}(0)$, then $x_{c_1}(t) < x_{c_2}(t)$ for all $t$. This enforces a stable column ordering and prevents overtaking. This constraint fundamentally shapes execution. Arbitrary permutations cannot be realized via lateral motion alone; instead, routing must use (1) monotone translations of multiple columns or (2) AOD$\leftrightarrow$SLM transfers (known as trap changes) to change column membership. As shown in Fig.~\ref{fig:aod}, even logically independent CZ operations can conflict under non-crossing, introducing serialization~\cite{barredo2016assembler,endres2016assembly,pachinqo2024}.

\paragraph{Rydberg interactions and blockade}
Two-qubit gates rely on the \emph{Rydberg blockade}, where excitation of one atom suppresses excitation of nearby atoms. Gates are only valid within a fixed interaction radius, which simultaneously defines an exclusion region: any unintended atom within this radius will be entangled. As a result, execution must carefully control both placement and movement to avoid interference~\cite{saffman2010rmp,weiss2017quantum}.

\subsection{Quantum Error Correction Basics}

Quantum error correction (QEC) encodes $k$ logical qubits into $n$ physical qubits with distance $d$, denoted $\llbracket n,k,d\rrbracket$, enabling detection and correction of errors during execution~\cite{terhal2015rmp,nielsen2010,xu2024constant}. In stabilizer codes, this is achieved by repeatedly measuring commuting parity checks (stabilizers), whose outcomes form a syndrome used to infer and correct errors. A widely used subclass is \emph{CSS codes} (Calderbank-Shor-Steane)~\cite{calderbank1996css,steane1996prl}, which separate checks into $X$-type and $Z$-type operators. This separation enables structured execution via alternating $X$ and $Z$ rounds, allowing shallow circuits and a high degree of parallelism.

From a systems perspective, two properties dominate execution cost: \emph{check weight} (number of qubits per check) and \emph{degree} (number of checks per qubit). \emph{Quantum LDPC} (qLDPC) codes maintain both as constants under scaling, enabling bounded-depth, parallelizable syndrome extraction. \emph{Bivariate Bicycle (BB)} codes are CSS qLDPC codes with fixed small check weight and a highly regular, translational structure. In this work, we focus on the \emph{gross} BB code $\llbracket 144,12,12\rrbracket$ primarily~\cite{yoder2025tourdegross,bravyi2024bb}. Its constant-weight checks and uniform layout align well with constraints on neutral-atom movement, while its instruction set provides a compact interface for compilation on this reconfigurable architecture.

\begin{figure}[t]
\centering
\includegraphics[width=0.98\linewidth]{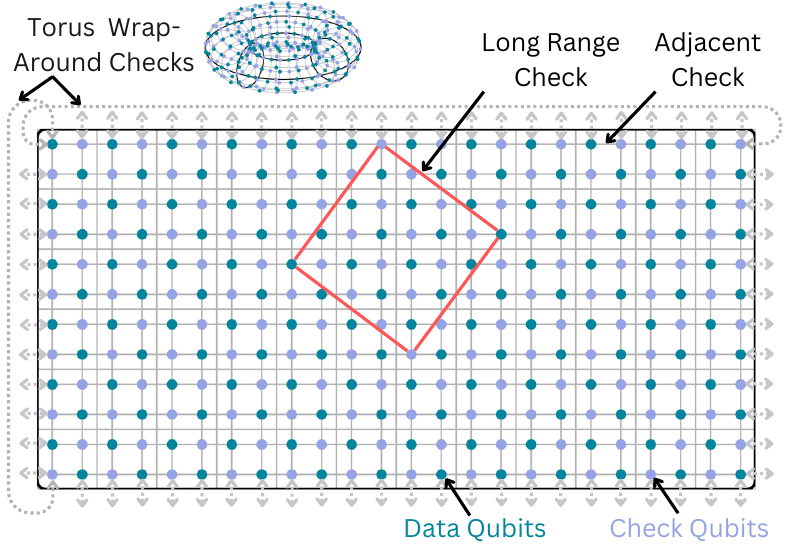}
\vspace{-2mm}
\caption{Module structure for BB codes. Note the long-range connections (only one set of edges displayed; these edges repeat across the torus). The wraparound edges capture the toroidal logical geometry.}
\vspace{-5mm}
\label{fig:module_struct}
\end{figure}

\subsection{BB Codes and Evaluated Gross-Code Instances}
\label{sec:bb-gross}

\paragraph{BB code construction}
BB codes can be viewed as a toroidal grid of identical \emph{unit cells}. Each cell contains two data qubits, \emph{left} ($L$) and \emph{right} ($R$), and participates in one $X$-type and one $Z$-type parity check. Each check has constant weight~6, and each data qubit participates in six checks (three $X$, three $Z$). Checks connect locally to neighboring cells and include two fixed long-range connections that wrap around the torus (Fig.~\ref{fig:module_struct}). For the $\llbracket 144,12,12\rrbracket$ gross code, an $\ell \times m = 12 \times 6$ grid yields $n=144$ data qubits and $k=12$ logical qubits with distance $d=12$. This regular structure enables a compact, repeatable execution pattern. The $\llbracket 288,12,18\rrbracket$ two-gross code extends this construction to a $12\times12$ grid, doubling the data- and check-qubit footprint while retaining the same BB-native execution interface.

\paragraph{Execution interface}
BB-code execution consists of:
\begin{enumerate}[leftmargin=15mm]
    \item \emph{Idle}: alternating $X$- and $Z$-syndrome rounds,
    \item \emph{Shift automorphisms}: structured permutations of the toroidal layout,
    \item \emph{In-module measurements}: local logical operations,
    \item \emph{Inter-module measurements}: joint operations between modules,
    \item \emph{$T$-injection}: magic-state injection on logical qubits.
\end{enumerate}

CSS codes do not require strict separation of $X$ and $Z$ checks; however, in our work, we separate them to align with zoned execution, where operations are batched into motion-compatible phases. Alternative interleavings correspond to different orderings of the same underlying transport, entangling, and measurement primitives.

%% file: sections/motivation.tex
\section{Related Work and Motivation}
\label{sec:motivation}

QEC has been extensively explored across superconducting, trapped-ion, photonic, and neutral-atom platforms, with progress spanning surface codes, color codes, small-distance LDPC codes, and lattice-surgery-based protocols~\cite{caliqec,xu2024constant,zhou2025lowoverheadtransversalfaulttolerance,Kim2024wob,salesrodriguez2025msd,wang2024optimizing}. Within this space, BB codes~\cite{bravyi2024bb} provide a promising class of constant-weight, constant-degree qLDPC codes enabling constant-depth stabilizer extraction. 

Prior work has focused on code construction, decoding, and logical instruction sets for BB and generalized bicycle codes, including the Tour de gross (Tdg) architecture~\cite{yoder2025tourdegross}, which defines a BB-native logical interface and compilation strategy. However, these works remain at the logical level or target alternative architectures, and do not provide a pathway to execute BB primitives under the physical constraints of neutral-atom systems. \sol{} complements prior BB architecture efforts that define instruction sets and logical compilation, often assuming modular machines with long-range couplers or Bell-pair links~\cite{yoder2025tourdegross,cross2025improved}. In contrast, \sol{} focuses on executing these BB-native abstractions on zoned neutral-atom arrays with SLM storage, AOD transport under non-crossing and stable-ordering constraints, and Rydberg blockade restrictions. Our contribution is to translate BB-native operations into legal neutral-atom motion schedules that preserve logical semantics while respecting these hardware constraints. Rather than introducing new BB fault-tolerant primitives, \sol{} provides the missing hardware-grounded execution layer needed to realize existing BB interfaces.

The recent two-stage pipeline of Sethi et al.~\cite{sethi2026logical} clusters logical qubits into modules and orders the resulting modules using abstract mapping costs; its only direct overlap with \sol{} is this module-ordering stage, which corresponds to our spectral placement, whereas \sol{} realizes module-level operations as legal placements and motion schedules on zoned neutral-atom hardware. The approaches are therefore composable---their clustering can feed \sol{}'s placement and scheduling, reducing the number of inter-module measurements while \sol{} reduces the cost of each remaining measurement, including 34\% lower bridge travel and 45\% fewer serialized bridge rounds (Sec.~\ref{sec:evaluation}).

In parallel, the neutral-atom community has developed scalable, reconfigurable qubit arrays using SLMs, AODs, and Rydberg-mediated entangling gates. Existing compilers, including Q-Pilot~\cite{wang2024q} and Atomique~\cite{wang2024atomique}, as well as reinforcement-learning-based mapping strategies~\cite{nakaji2025} and timing-aware routing tools~\cite{Schmid_2024,stade2025,Wille_2023}, focus on physical-level circuits or surface-code-like layouts. These approaches do not support BB-native primitives, qLDPC-specific module structures, or bridge-driven logical operations. Similarly, methods that combine generalized-bicycle (GB) codes with separate computational layers~\cite{viszlai2024} lack mechanisms to execute BB logic on neutral-atom architectures. Despite progress in both BB-code design and neutral-atom compilation, there remains no hardware-aware execution model, placement strategy, or movement-aware compiler capable of realizing BB logical instructions under SLM, AOD, and Rydberg-blockade constraints. This work addresses that gap by introducing a constraint-compatible execution model that enables BB-native operations to run directly on neutral-atom systems.

%% file: sections/design.tex
\section{\sol{}'s Design}
\label{sec:design}

We describe how a BB-native circuit is compiled into neutral-atom hardware actions in \sol{}. The goal is to translate logical BB operations into a hardware-executable schedule comprising shift automorphisms, idling cycles, and intra- and inter-module measurements within a tiled compute zone composed of compute columns, subzones, logical processing units (LPUs), and bridge qubits.

A BB-native circuit is expressed over modules, stabilizer measurements, and automorphisms that permute atoms or rotate the lattice. \sol{} maps this circuit to hardware by determining module placement, grouping modules to minimize movement, interleaving stabilizer cycles with shift automorphisms, and realizing joint measurements via bridge-qubit transport while respecting AOD constraints.

BB-native execution is governed by three constraints. First, BB codes require structured long-range interactions and toroidal wrap-arounds, which must be implemented through physical transport. Second, AOD transport enforces non-crossing and stable ordering. Third, inter-module measurements require coordinated bridge motion without violating the blockade or introducing collisions. \sol{} addresses these through (1) spectral placement to co-locate frequently interacting modules and reduce transport cost, (2) a shift-automorphism decomposition into directional rolls with resynchronization that preserves ordering, and (3) an interval-based bridge scheduler that guarantees conflict-free execution under zoning, blockade, and AOD constraints.

The compilation pipeline consists of three stages: (1) logical layerization into parallelizable layers and sublayers, (2) spatial placement of modules into compute columns, and (3) hardware scheduling that converts each layer into column- and subzone-level actions for shifts, measurements, and idle cycles. We describe each stage below.

\begin{figure}[t]
\centering
\includegraphics[width=0.99\linewidth]{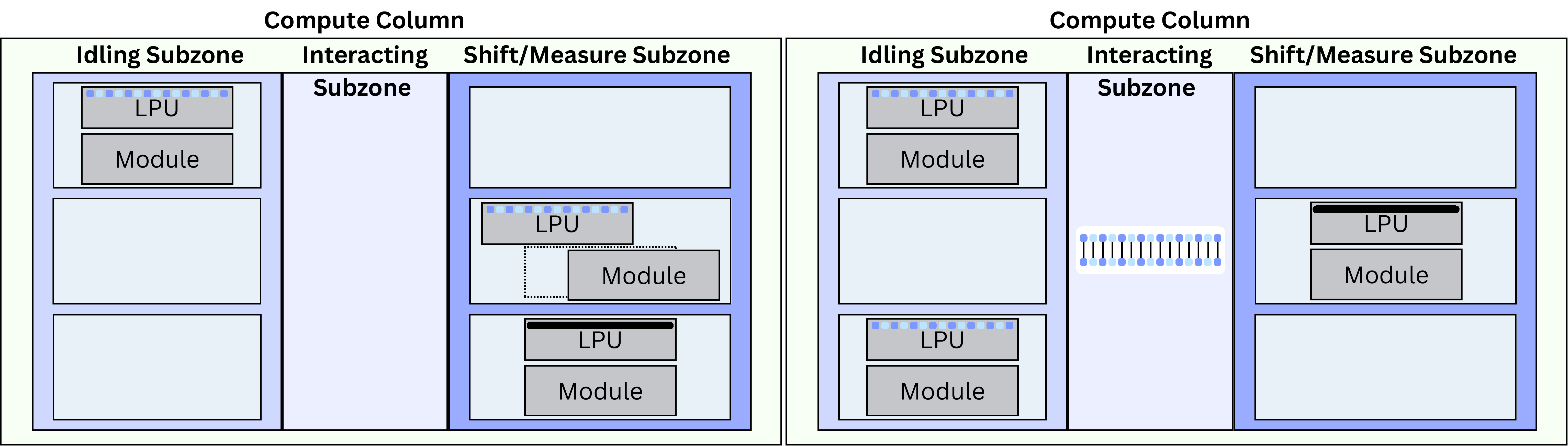}
\vspace{-5mm}
\caption{Compute zone with two columns (capacity 3 modules each). Modules execute idle, shift, and joint-measurement operations across idling, interaction, and shift/measurement subzones. Bridge qubits from interacting modules are brought together in the interaction subzone. Shift and measurement occur in separate sublayers but are co-visualized.}
\vspace{-5mm}
\label{fig:overview}
\end{figure}

\subsection{Layerization of the BB Native Circuit}

\paragraph{Layering of the Circuit}
The front end represents a BB-native circuit as a tableau, where each entry corresponds to an automorphism, a single-module measurement, or a two-module joint measurement over module indices. This representation is purely logical and does not encode physical placement. We construct a dependency-directed acyclic graph (DAG)~\cite{JavadiAbhari_2015,Sivarajah_2020} where each node corresponds to a BB operation, and edges capture per-module ordering constraints. Operations acting on disjoint modules remain unordered, exposing available parallelism. From this DAG, we derive a maximally parallel layered schedule via topological ordering, grouping operations into layers such that (1) all dependencies are satisfied and (2) no two operations in the same layer act on the same module. Each layer represents a logical timestep with maximal concurrency. Layers are refined into sublayers aligned with execution semantics. Measurement-only or idle-only layers collapse to a single sublayer, while layers containing both shifts and measurements are split into two ordered sublayers. Transitions between sublayers relocate modules across subzones as needed. The output of this stage is a per-timestep sequence of column-local actions (shift, measure, idle) that preserves logical dependencies while preparing for constraint-aware hardware execution.

\paragraph{Column Construction}
We map the logical schedule to hardware by partitioning the compute zone into vertical \textit{compute columns}, each hosting a fixed stack of BB modules. Each column is divided into three subzones: (1) an \textit{idling} subzone for error-checking cycles, (2) a \textit{bridge-interaction} subzone for inter-module operations via bridge qubits, and (3) a \textit{shift/measurement} subzone where all active operations are executed under column-synchronous Rydberg pulses.

A logical sublayer is realized by moving only the participating modules (and their bridge rows when needed) into the appropriate active subzone, while all other modules remain idle. This column-local organization enables hardware-wide synchronous execution while confining motion within columns, largely avoiding violations of AOD non-crossing constraints of the system (Fig.~\ref{fig:overview}).

\subsection{Compute Zone Construction and the LPU}

Joint measurements require entangling qubits from two modules via \textit{bridge qubits}, which interact in the bridge-interaction subzone. Each module is associated with a \textit{logical processing unit (LPU)} positioned above it, which houses these bridge qubits. The LPU layout is designed to simplify joint measurements. The top row contains bridge and bridge-check qubits arranged to enable efficient Bell-pair generation, which underlies inter-module measurements. Lower rows host ancilla qubits for stabilizer measurements. Joint operations are executed by transporting bridge qubits from participating modules into the interaction subzone for proximity.

A key challenge is that distant modules require long bridge movements, which are costly under AOD non-crossing constraints. To mitigate this, \sol{} employs a spectral placement strategy that assigns modules to compute columns based on an interaction graph, placing frequently interacting modules close together and reducing bridge movement.

\begin{figure}[t]
    \centering
    \includegraphics[width=0.99\columnwidth]{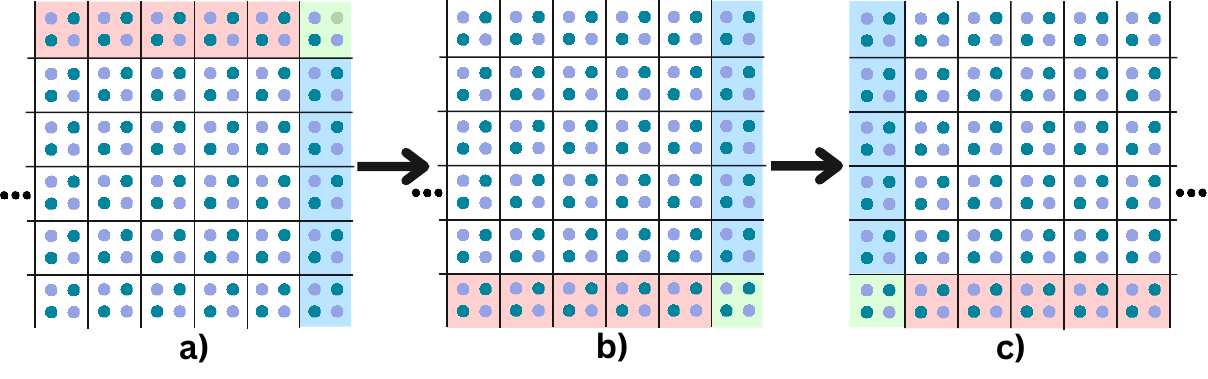}
    \vspace{-6mm}
    \caption{Decomposition of $T(\Delta i,\Delta j)$ with $\Delta i,\Delta j=+1$: (b) vertical roll $\mathrm{Roll}_i$ (top row to bottom), followed by (c) horizontal roll $\mathrm{Roll}_j$ (rightmost col. to left), applied to (a).}
    \vspace{-5mm}
    \label{fig:equivalentshifts}
\end{figure}

\subsection{Idling and Error-checking During Phases}
\label{subsec:idling}

In the bicycle architecture~\cite{yoder2025tourdegross,bravyi2024bb}, an \emph{idle} on a module corresponds to one full round of stabilizer measurements, where all $X$- and $Z$-type checks are evaluated, and the net logical operation is the identity. \sol{} preserves this behavior but must realize it under AOD movement constraints and toroidal BB connectivity (Fig.~\ref{fig:module_struct}). Execution is organized at the sublayer level. Modules in the \emph{idling subzone} continuously perform error checking, executing a full stabilizer schedule in every sublayer in which they remain idle. Modules in the \emph{shift/measurement subzone} do not error check during that sublayer to avoid conflicts in AOD motion. To maintain a consistent notion of logical time, idling modules execute the same number of error-checking cycles as active modules (ten cycles per measurement sublayer~\cite{yoder2025tourdegross}).

At the end of each logical layer, \sol{} inserts a global idling sublayer in which all modules execute a synchronized, wrap-aware stabilizer cycle. This enforces a consistent error-correction cadence across the compute zone.

The idling scheduler implements BB stabilizer extraction as a fixed 24-substep sequence of AOD movements and entangling operations. Each substep moves ancilla qubits along a single monotone direction and groups interactions by wrap class on the torus, ensuring that all stabilizers are measured exactly once while preserving non-crossing, column-ordered motion. Adjacent and long-range checks are partitioned into a small number of direction- and wrap-consistent groups, enabling conflict-free execution under AOD constraints. This structured schedule maintains BB error protection while minimizing movement overhead and avoiding ordering violations. \sol{} records idling gate counts and AOD travel distances across modules and sublayers, enabling architecture-level estimation of error-protection cost (used in Sec.~\ref{sec:evaluation}).

\subsection{Shift-Automorphism Scheduling in the Compute Zone}
\label{sec:shiftauto}

We formalize how to schedule shift automorphisms across modules in a neutral-atom compute zone under AOD constraints. The key observation is that toroidal shifts are direction-agnostic: each shift can be realized via complementary edge movements, enabling parallel, column-aligned execution. A shift $(\Delta i,\Delta j)$ on an $N_i\times N_j$ torus decomposes as $T(\Delta i,\Delta j)=\mathrm{Roll}_i(\Delta i)\circ \mathrm{Roll}_j(\Delta j)$, where $\mathrm{Roll}_x(\Delta)$ cyclically shifts along axis $x$. By periodicity, $\mathrm{Roll}_x(\Delta)=\mathrm{Roll}_x(\Delta-N_x)$ and $\mathrm{Roll}_x(-\Delta)=\mathrm{Roll}_x(N_x-\Delta)$, so a roll can be implemented in either direction. Let $k_i\equiv \Delta i \bmod N_i$ and $k_j\equiv \Delta j \bmod N_j$. The horizontal roll moves either a width-$k_i$ chunk right or a width-$(N_i-k_i)$ chunk left; the vertical roll similarly moves up or down. These complementary moves yield the same logical translation (up to a global offset corrected later), giving four equivalent realizations per shift.

A shift is executed as two rolls (along $i$ and $j$) followed by a lightweight resynchronization. Each roll moves only the selected wrap-around strips within the shift/measurement subzone and returns them to their lattice positions. Because modules may have different shift magnitudes, the two rolls introduce small horizontal and vertical offsets within a column. \sol{} corrects these via a monotone resynchronization sweep, where modules move in a common direction and drop to SLM once aligned.

All steps respect AOD non-crossing constraints: AOD columns are monotonically ordered and cannot overtake one another. \sol{} enforces legality by (i) using consistent physical directions per column during shifts, ensuring stable-ordered motion, and (ii) performing resynchronization via a monotone sweep with staggered SLM dropoffs. This preserves ordering while exposing when parallel actions must serialize.

Given a parallel layer $L$ of shifts partitioned into compute columns, \sol{} executes two AOD phases per layer: one for horizontal rolls and one for vertical rolls (Fig.~\ref{fig:equivalentshifts}). Each column elects a head module (e.g., the top-most) with shift $(\Delta i^\star,\Delta j^\star)$. The column adopts the head’s physical directions on both axes. Every module executes its shift using this shared convention, selecting its own rows/columns but moving them toward the head. Because all modules share directions, motion is monotone and can be executed in parallel without AOD crossings. This parallelism introduces residual offsets from differing shift magnitudes, reducing future parallelism. We address these via a resynchronization policy described next.

\begin{figure}[t]
    \centering
    \includegraphics[width=0.99\columnwidth]{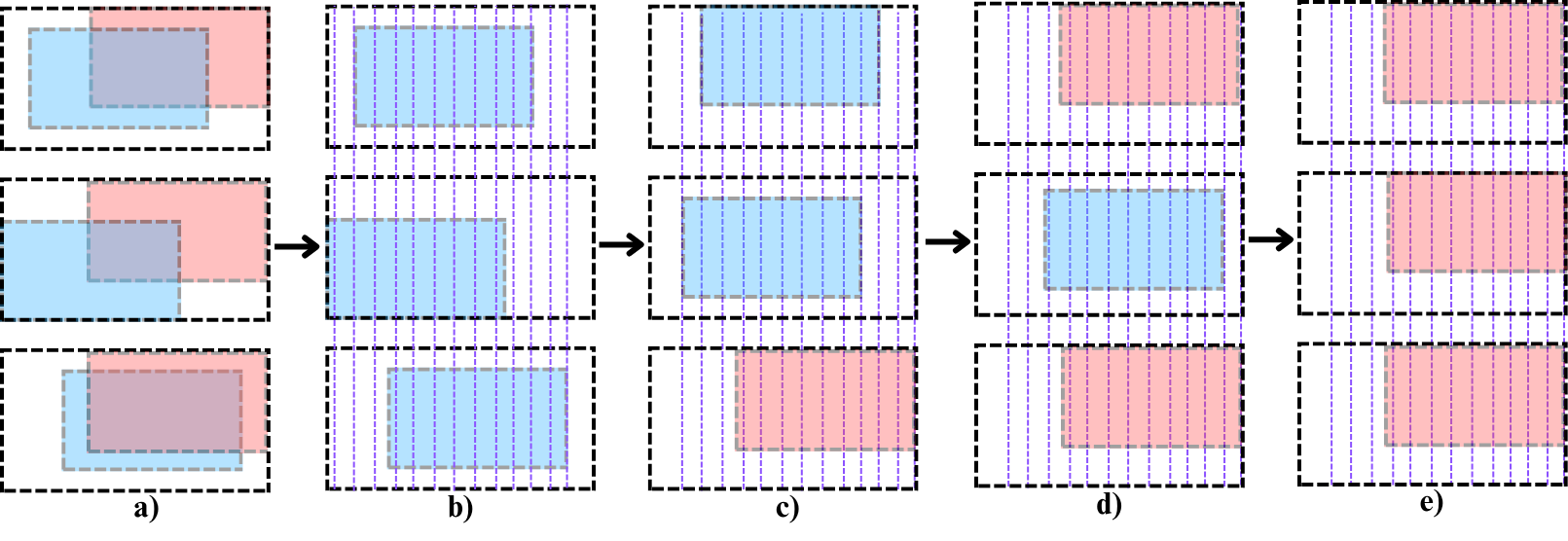}
    \vspace{-6mm}
    \caption{\textbf{a)} After a shift, modules (blue) are offset and must be returned to their original locations (\textit{resync}). \textbf{b)} All modules are captured by shared AOD traps and moved in parallel. \textbf{c)} Modules are restored in order of increasing distance; all move 2 units right and 2 units up, after which the bottom module reaches its target and drops to SLM. \textbf{d)} The process repeats for the remaining modules; the middle module waits while the top module drops. \textbf{e)} The final module is moved and dropped; all modules are resynchronized.}
    \vspace{-5mm}
    \label{fig:resync}
\end{figure}

\subsubsection{Residual Misalignment and Why Magnitude Diversity Causes Collisions}

After the two wraps, modules in a column differ in absolute position due to heterogeneous shift magnitudes (e.g., some move 2 columns, others 5) in the same direction. Independently correcting each module would require longer moves to pass shorter ones, violating the AOD no-crossing constraint. 

\sol{} instead performs a monotone resynchronization sweep: all modules move in a common direction and drop off once they reach their targets. Let $r_{(m)}^i\ge 0$ and $r_{(m)}^j\ge 0$ denote the residual horizontal and vertical distances to the chosen alignment (e.g., left/top edges flush). The resync direction is chosen so all residuals are nonnegative (matching the head’s direction), ensuring feasible motion.

Sorting horizontal residuals yields $r_{(1)}\le r_{(2)}\le \cdots \le r_{(M)}$, with increments
\begin{equation}
\Delta_1=r_{(1)},\qquad
\Delta_k=r_{(k)}-r_{(k-1)}\quad\text{for }k\ge 2.\notag
\end{equation}

At each step, all active modules are translated by $\Delta_k$. Modules that reach their targets drop to SLM, while the rest continue. Because motion is monotone, module ordering is preserved, and no crossings occur. The number of moves equals the number of distinct residual magnitudes, bounded by the number of modules. The vertical sweep is identical, and both axes can be executed concurrently without violating constraints (Fig.~\ref{fig:resync}).

Putting this together, for each parallel shift layer, each compute column selects a head, fixes directions, executes the two wraps in parallel, and performs resynchronization via staggered drop-offs. Columns operate independently, and per-column latency is the cost of two rolls plus a small number of resync steps. Correctness follows from torus periodicity and the preservation of AOD ordering. Any torus shift can be realized via minimal or complementary edge movements, chosen independently per axis and module. Thus, \sol{} supports arbitrary mixes of shifts within a column. \textit{The resulting schedule is direction-agnostic and directly compatible with AOD movement and Rydberg constraints.}

\begin{figure*}[t]
    \centering
    \includegraphics[width=0.99\textwidth]{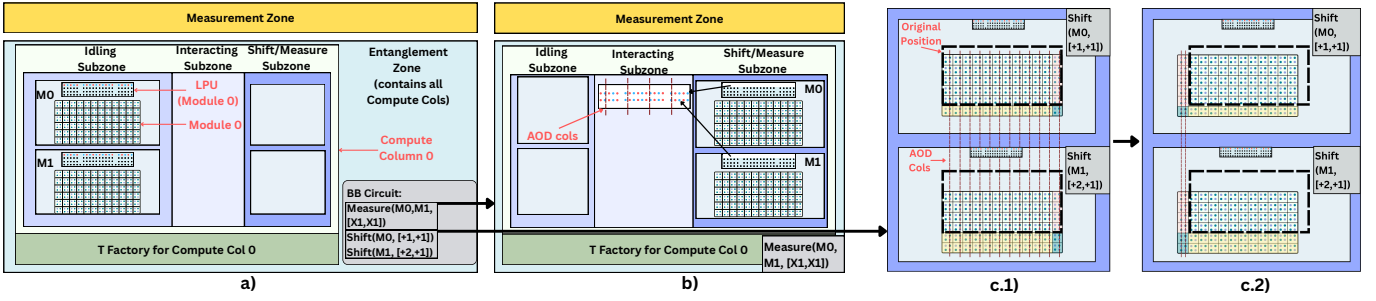}
    \vspace{-1mm}
  \caption{End-to-end example of a 2-module, single compute-column execution. (a) Modules 0 and 1 are placed in the idling subzone via spectral placement and remain within the column, moving only between subzones. (b) A joint measurement: modules move to the shift/measurement subzone, bridge qubits are brought within interaction range via AOD to form Bell pairs, then return to LPUs. (c.1) Parallel shifts $(+1,+1)$ (Module 0) and $(+2,+1)$ (Module 1) using shared AOD directions, producing a vertical offset. (c.2) Horizontal rolls complete the shift; resynchronization (Fig.~\ref{fig:resync}) restores alignment.}
  \label{fig:worked_ex}
  \vspace{-5mm}
\end{figure*}

Finally, Fig.~\ref{fig:worked_ex} illustrates a worked example with a full layer: modules are placed across subzones, shifts execute as two ordered rolls followed by resync, and joint measurements are performed via bridge motion. Each micro-step is either a monotone AOD translation or an AOD/SLM transfer, ensuring legal execution under the non-crossing constraint.

\subsection{Static Module Placement via Spectral Seriation}
\label{sec:spectral_placement}

To perform joint measurements, bridge qubits from interacting modules must be brought into close proximity. Since AOD columns can be translated but not crossed, module placement within compute columns directly affects communication cost. To minimize bridge movement, modules that frequently interact (i.e., share many joint measurements) should be co-located. We address this via a graph-based spectral seriation algorithm~\cite{fogel2016spectralrankingusingseriation,acharya2022qubitseriation}. The procedure has three steps.

(1) \textit{Interaction Graph Construction:} We model the circuit as a weighted, undirected graph $G=(V,E)$, where each vertex represents a module and edge weights $W_{ij}$ count how often modules $i$ and $j$ participate in joint measurements within the same circuit moment. This yields a symmetric adjacency matrix $W\in\mathbb{R}^{n\times n}$ with $n=|V|$.

(2) \textit{Spectral Seriation:} We compute the graph Laplacian $L=D-W$, where $D_{ii}=\sum_j W_{ij}$. The Fiedler vector (eigenvector corresponding to the second-smallest eigenvalue of $L$) provides a 1D embedding that places strongly interacting modules close together. Sorting modules by their Fiedler coordinates produces an ordered list.

(3) \textit{Column Packing:} We partition this ordering into contiguous blocks based on hardware column capacity, assigning modules sequentially to compute columns.

This approach co-locates strongly interacting modules within the same or nearby columns, directly reducing inter-column communication under AOD constraints.

\subsection{Bridge-based Joint-measurement Scheduling}
\label{sec:bridge-scheduler}

Inter-module joint measurements are realized by moving a dedicated bridge row of ancilla atoms from each module’s LPU into the interaction subzone of a compute column, where Rydberg pulses entangle them into Bell pairs. Given a logical layer $\ell$ with fixed module placement and a set of joint-measurement requests $\{(m_i,m_j)\}$ (each module participates in at most one), \sol{} converts these requests into a sequence of bridge moves that obey AOD no-crossing constraints.

We model the interaction lane as a 1D array of columns $C=\{0,\dots,C{-}1\}$ and assign each module $m$ a home column $h_m\in C$. For a pair $(m_i,m_j)$, we define the interval $I_{ij}=[\min(h_{m_i},h_{m_j}),\max(h_{m_i},h_{m_j})]$, representing the corridor within which the two bridges move. Pairs are partitioned into \emph{rounds} via greedy interval coloring so that intervals within a round are disjoint. As a result, no two joint measurements in a round compete for the same column.

For each pair, we select a meeting column $c_{ij}$ as the midpoint of $I_{ij}$. Starting from $c^{(0)}_m=h_m$, the scheduler advances in discrete micro-steps $k=1,2,\dots$. At each step, each participating bridge moves by at most one column toward $c_{ij}$ (left, right, or stay). The resulting positions $\{m\mapsto c^{(k)}_m\}$ define a movement layer $L_k$. Because intervals are disjoint and $c_{ij}\in I_{ij}$, these lockstep $\pm1$ updates preserve column ordering and avoid collisions.

Once all bridges reach their meeting columns, joint measurements are executed in parallel. A symmetric reverse sequence returns all bridges to their home columns. Overall, the bridge scheduler maps joint-measurement requests to a column-wise movement plan. \textit{Disjoint intervals ensure non-interfering groups, and unit-distance updates guarantee physically realizable, non-crossing AOD motion.}

\subsection{Per-column T-state Factories}

\sol{} targets the hardware-realizable execution of BB-code primitives (shifts, in-module measurements, and inter-module joint measurements) under neutral-atom motion and zoning constraints. Non-Clifford operations are supplied via magic-state distillation (MSD) and injection; the specific MSD protocols and any code-conversion or lattice-surgery procedures are orthogonal to \sol{}. However, MSD has a concrete architectural footprint: it consumes dedicated area, requires a data/factory interface, and introduces a throughput constraint. To capture this, we augment each compute column with a \emph{factory area} beneath the compute zone. Each column is provisioned with one colocated T-state factory, and injection is performed locally (e.g., by moving the target module to a local interface region and applying an injection sequence). This ``one factory per column’’ organization follows \sol{}’s core design principle: constrain long-range communication to joint measurements while keeping high-rate, latency-sensitive services local to each column.

In our evaluation (Sec.~\ref{sec:evaluation}), we model each factory as a single-server resource with service time $\tau_{\mathrm{fac}}$ per magic state, and each injection incurs a non-overlappable local overhead $\tau_{\mathrm{inj}}$. This protocol-agnostic abstraction allows any MSD approach, characterized by per-column throughput and local consumption overhead, to be integrated. We use this model to identify when factory throughput becomes a bottleneck.

Next, we summarize the design of \sol{}.

\subsection{Putting it All Together}

\sol{} presents a hardware-aware execution model that co-designs BB-code primitives with neutral-atom constraints. Spectral placement reduces communication distance, shift-automorphism scheduling exploits direction-agnostic torus structure for parallel execution, and bridge-based scheduling enables conflict-free inter-module measurements.

These components operate within a columnar organization that localizes motion, respects AOD non-crossing constraints, and exposes structured parallelism. Finally, per-column T-state factories integrate non-Clifford execution into this framework without introducing global communication bottlenecks. Together, these techniques provide a practical path from BB-code abstractions to high-throughput execution on reconfigurable neutral-atom systems.

%% file: sections/methodology.tex
\section{Experimental Methods and Implementation}
\label{sec:methods}

We evaluate \sol{} on 30 BB circuits with an average joint-measurement correlation of $0.35$ (higher correlation means that each module tends to interact with a smaller set of partner modules). The circuits evaluated range in size from 32 to 113 modules, each comprising 144 physical data qubits and 144 ancilla check qubits (288 physical qubits per module), as well as 12 logical qubits (the $\llbracket144,12,12\rrbracket$ gross BB code). Thus, these modules range from 9,216 to 32,544 physical qubits and 384 to 1,356 logical qubits. We also evaluate the circuits using the $\llbracket288,12,18\rrbracket$ two-gross BB code for scalability. Each circuit applies 10 logical operations per module, excluding background idle error-checking, and $25\%$ of all non-idle operations are joint measurements. Our experiments are implemented in Python~3.11.9, and all reported runtimes in this section are obtained on a standard Apple computer with an M1 chip, 8~CPU cores, and 8~GB RAM.

\paragraph{Neutral Atom Hardware Model}
We adopt a neutral atom hardware model consistent with prior work on compilers for these devices~\cite{ludmir2024parallax,pachinqo2024,bluvstein2024logical}. In this model, AOD shuttling occurs at a speed of $v = 55~\mu\mathrm{m}/\mu\mathrm{s}$, with a Rydberg interaction distance of  $2~\mu\mathrm{m}$ and a lattice separation distance between physical qubits in a module of $10~\mu\mathrm{m}$ to avoid crosstalk. All Rydberg pulses apply in a global fashion to the qubits in the compute zone~\cite{bluvstein2024logical}. Two-qubit and single-qubit gate durations are fixed to $0.8~\mu\mathrm{s}$ and $2.0~\mu\mathrm{s}$ respectively, and we model the cost of CXs as a $\mathrm{CZ}$ plus two single-qubit pulses. Each pickup or dropoff between SLM and AOD traps incurs a trap-switch latency of $100~\mu\mathrm{s}$. To isolate the impact of layout and shuttling (which is what we optimize for), we set the measurement time to a negligible value and model-check the qubit reset to the $\ket{+}$ state as a single-qubit pulse. The reason for this is that if we were to use long measurement times from current hardware, such as Aquila~\cite{wurtz2023aquila}, where a full measurement cycle takes $\mathcal{O}(10^2)$~ms, these values would dominate our runtime metric and obscure differences between compilation strategies.

We use runtime to evaluate how much transport and congestion-induced latency each compilation strategy adds. Because the BB circuit semantics are held fixed across strategies, the differences we report primarily reflect shuttling distance, trap-switching overheads, and AOD-induced serialization. This complements code-theoretic metrics such as thresholds and asymptotic logical-error scaling, which depend on the decoder choice and a detailed noise model beyond the scope of this systems paper. If each syndrome/measurement cycle incurs a fixed readout-and-classical-latency cost $T_{\mathrm{meas}}$ that is common across compilation strategies, then total schedule time becomes $T_{\mathrm{tot}} = T_{\mathrm{sched}} + N_{\mathrm{cyc}} T_{\mathrm{meas}}$, where $T_{\mathrm{sched}}$ is the transport/gate schedule time we model and $N_{\mathrm{cyc}}$ is the number of required cycles. In this regime, the absolute runtime savings and ordering between strategies are preserved. If readout overlaps with transport, as is typical in pipelined zoned execution, then reducing transport directly improves the critical path~\cite{wang2024q}. However, the dominant effect of Park-n-Ride relative to our baselines is to reduce total shuttling distance and the number of SLM$\leftrightarrow$AOD handoffs; under standard per-distance or per-move loss models, these reductions would be expected to decrease exposure to loss events rather than increase it. Atom loss would not impact the qualitative ordering between the competitive strategies.

On the atom array simulation side, each BB module is mapped to a rectangular footprint of $12\times 6$ lattice cells, so with the aforementioned spacing between qubits of $ 10~\mu\mathrm{m}$ the module width and height are $W_{\mathrm{mod}} = 120~\mu\mathrm{m}$ and $H_{\mathrm{mod}} = 60~\mu\mathrm{m}$. Compute columns stack these modules vertically and place their respective LPUs directly above each module. Horizontally, each compute column is partitioned into three subzones (idle, interact, shift/measure). The idle subzone is $1.2W_{\mathrm{mod}}$ in width to ensure space to avoid crosstalk with neighboring compute columns; the interact subzone is $1.1W_{\mathrm{mod}}$ since fewer qubits interact here but we still need room for the bridge qubits; the shift/measure is significantly larger at  $2W_{\mathrm{mod}}$ because the furthest distance modules can move during a shift automorphism is $0.5W_{\mathrm{mod}}$ in width, meaning this extra space is necessary to ensure modules do not exit the compute zone during these shifts. A full compute column thus occupies $4.3W_{\mathrm{mod}}$ in horizontal extent.

Given a compiled circuit, we estimate its runtime by summing per-layer contributions over all logical layers. For each layer, we compute (1) shuttling time from all AOD moves as (distance$/v$) plus the appropriate trap-switch overheads, (2) gate time from the counts of $\mathrm{CZ}$/$\mathrm{CX}$ and single-qubit pulses in the BB stabilizer gadgets and logical measurements, and (3) lane/subzone-transition time from any global subzone moves, modeled as a single parallel translation of all active modules. The \sol{} compiler outputs per-layer movement and operation statistics, and our estimator plugs those counts into this timing model to obtain $T_{\mathrm{est}}$, the estimated runtime of the circuit in milliseconds. Since virtually every important metric is highly correlated with the overall runtime of the circuit (e.g., parallelism, decoherence over time, good module placement), we consider this the most important end-performance metric for evaluation.

\paragraph{Readout-Time and Atom-Loss Sensitivity} We post-process the same 150 compiled schedules for each placement strategy (30 circuits across five compute-column capacities) to evaluate both sensitivities without recompilation. For end-to-end timing, we evaluate every schedule at \(T_{\mathrm{meas}}=0\) and \(1~\mu\mathrm{s}\) to extract its affine dependence \(T_{\mathrm{tot}}(T_{\mathrm{meas}})=T_{\mathrm{tot}}(0)+N_{\mathrm{meas}}T_{\mathrm{meas}}\), and then sweep \(T_{\mathrm{meas}}\in\{10,50,100,500,1000,1500\}~\mu\mathrm{s}\); setting \(T_{\mathrm{meas}}=0\) removes only the readout pulse time, while transport, trap handoffs, reset, and gate times remain included. For atom loss, a post-processing pass aggregates each schedule's total AOD shuttling distance \(D\) (in \(\mu\mathrm{m}\)) and number of SLM--AOD trap transfers \(S\) from measurement motion, bridge moves, shift automorphisms, and subzone transitions~\cite{kobayashi2026erasure,chow2024circuit,perrin2026correlated}. Under independent per-distance and per-transfer loss rates \(p_d\) and \(p_s\), respectively, the probability of at least one transport-induced loss is \(p_{\mathrm{loss}}=1-(1-p_d)^D(1-p_s)^S\), with first-order expected loss exposure \(\lambda\approx p_dD+p_sS\). We evaluate the \(4\times4\) logarithmic grid \(p_d\in\{10^{-8},10^{-7},10^{-6},10^{-5}\}~\mu\mathrm{m}^{-1}\) and \(p_s\in\{10^{-6},10^{-5},10^{-4},10^{-3}\}\) per transfer and compare the placement strategies at every grid point.

\paragraph{Comparative Techniques}
Due to the lack of prior work in this area, there are no state-of-the-art techniques to compare against. Thus, we examine several options for determining the optimal module orders in the compute columns. In addition to the spectral placement algorithm discussed earlier, we evaluate a \textit{hub-centric greedy heuristic} that directly constructs a low-cost 1D module ordering. As in the spectral method, we first build the weighted interaction matrix $W$ over BB modules. Given the number of modules and the per-column capacity, we determine the number of compute columns and select a set of ``hub'' modules with the largest interaction degree. We assign one hub to each column and seed them in a center-out pattern, so that the most highly connected modules occupy the central columns. The remaining modules are then placed one by one in a greedy fashion: at each step, for every unplaced module $u$ and every column $c$ that is not yet full, we evaluate a cost $\text{Cost}(u, c) = \sum_{v \in \text{PlacedNbrs}} W_{uv} \, |\mathrm{col}(c) - \mathrm{col}(v)|$, which measures the weighted 1D interaction distance to already placed neighbors, along with small penalties that reserve some room near each hub and break ties, choosing the assignment with minimum cost. As a final baseline alternative, we test an \textit{arbitrary method} in which modules are simply placed in the first available compute column. 

\paragraph{Non-Clifford Factory Sensitivity Model} We model each compute column as having one colocated T-state factory that serves all modules assigned to that column. Each factory is a deterministic single-server with service time $\tau_{\mathrm{fac}}$ per produced magic state. An injection request additionally incurs a non-overlappable local overhead $\tau_{\mathrm{inj}}$, which captures local rendezvous cost and state consumption. Starting from the same compiled schedules used in our main evaluation, we insert $k_T$ synthetic T-injection requests per module into the in-module measurement portion of the schedule. This isolates incremental factory cost while preserving Park-n-Ride’s baseline motion plan. We simulate factory cost with a per-column deterministic queue, where a request arriving at time $t$ begins service at $\max(t, t_{\mathrm{free}})$ and updates the column’s factory-available time $t_{\mathrm{free}}$. Because measurement and other phase work can overlap with waiting under our abstraction, we report factory impact primarily as the worst-case factory stall, defined as the maximum per-injection queue wait within a run. We also track the unavoidable local injection work $N_T(\tau_{\mathrm{inj}}+\tau_{\mathrm{fac}})$.

\paragraph{Evaluation Metrics} We evaluate the circuit \textit{compilation times, run times, and logical error rates}. To quantify how quickly shift automorphisms are consumed, we build cumulative progress curves from the execution traces. Each shift is logged as an event $e$ with layer index $\ell(e)$ and compute column index $c(e)$. In the parallel schedule (\sol{}'s technique), all shifts in the same layer $\ell$ execute concurrently, so we define $F_{\mathrm{par}}(L) = \frac{1}{N}\sum_{e} \mathbf{1}\!\bigl[\ell(e) \le L\bigr]$, the fraction of the $N$ shifts completed by layer $L$. We compare the \sol{} parallelized shift automorphisms with a serialized baseline, where we preserve inter–column parallelism but unroll intra–column parallelism such that for each layer $\ell$ and column $c$ we count $n(\ell,c)$ shifts, set $K_\ell = \max_c n(\ell,c)$, and expand layer $\ell$ into $K_\ell$ sublayers, each executing at most one shift per column. This assigns each event an effective serialized layer index $L'(e)$ and yields an analogous $F_{\mathrm{ser}}(L')$. For each circuit size, we average $F_{\mathrm{par}}$ and $F_{\mathrm{ser}}$ over all circuits, and plot the resulting curves as the cumulative fraction of shifts executed versus (original or serialized) layer index.

Similarly, to quantify bridge-movement parallelism, we build cumulative progress curves over the micro-steps of the bridge moves. Each entry in our bridge schedule corresponds to a micro-step  $t$ during which at least one logical bridge row is moved, and logs the total number of physical qubits transported in the sublayer. We infer the number of bridges moved in step $t$ as $B_t = \textit{qubit\_pickups}/q$, where $q$ is an approximate ``qubits-per-bridge'' value given by the greatest common divisor of all non-zero \textit{qubit\_pickups} for that circuit, and define $N = \sum_t B_t$. The parallel curve is $ F_{\mathrm{par}}(T) = \frac{1}{N} \sum_{t \le T} B_t$, the fraction of all bridge moves completed by micro-step index $T$ under the actual parallel schedule. For a serialized baseline, we conceptually unroll intra-step parallelism, replacing a step with $B_t$ bridges with $B_t$ unit steps, each moving a single bridge. We group circuits by size, average these curves within each group, and plot the resulting cumulative fractions versus the micro-step indices. Because these indices count finer-grained bridge-scheduler steps rather than logical layers, the horizontal axis is in the order of thousands.

%% file: sections/evaluation.tex
\section{Evaluation, Analysis, and Discussion}
\label{sec:evaluation}

\begin{figure}[t]
    \centering
    \includegraphics[width=0.99\columnwidth]{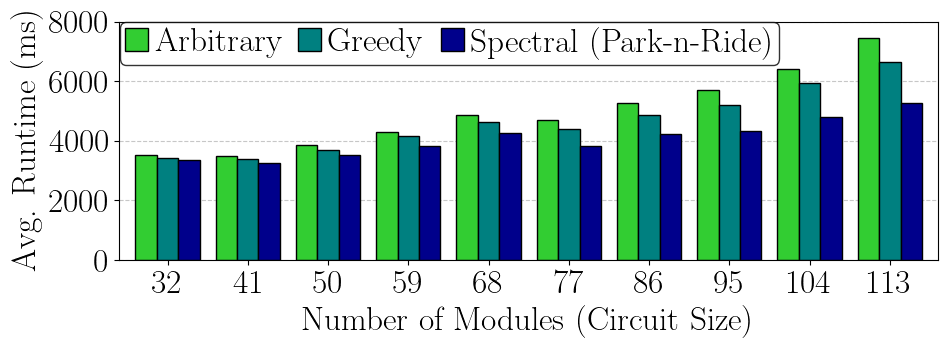}
    \vspace{-6mm}
    \caption{Comparison of average estimated runtime comparing module count between the three module placement techniques.}
    \vspace{-4mm}
    \label{fig:by_circuit_size}
\end{figure}

\paragraph{Flagship Results} Fig.~\ref{fig:by_circuit_size} compares the estimated runtimes for the three placement strategies as a function of circuit size, averaging over column capacities. For every size, the spectral method achieves the lowest estimated runtime, the arbitrary baseline the highest, and the greedy heuristic lies in between. The separation between the curves is modest at small module counts but grows steadily with size: at 113 modules, spectral placement is more than 2\,s faster than arbitrary placement, with greedy recovering roughly half of that benefit. This trend confirms that exploiting interaction structure to co-locate heavily connected modules becomes increasingly important as the BB graph grows, and that global structure (spectral) yields more benefit than purely local greedy decisions.

\begin{figure}[t]
    \centering
    \includegraphics[width=0.99\columnwidth]{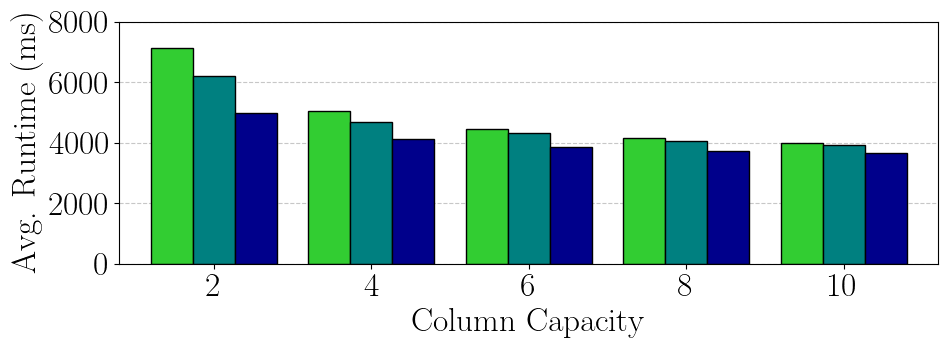}
    \vspace{-6mm}
    \caption{Comparison of the estimated runtime of three module placement techniques grouped by column capacity.}
    \vspace{-5mm}
    \label{fig:by_col_cap}
\end{figure}

Fig.~\ref{fig:by_col_cap} aggregates estimated runtimes of the three strategies by column capacity, averaging over circuit sizes. Increasing the capacity strictly improves performance for all placers, but with diminishing returns beyond the compute column capacity of 6. The largest improvement comes from avoiding very short columns; increasing the capacity from 2 to 4 reduces the average estimated runtime by roughly 30\% across all three placers. Interestingly, spectral placement with a capacity of 6 already outperforms arbitrary placement even at a capacity of 10, indicating that intelligent module placement can partially substitute for additional hardware resources. These results highlight a co-design trade-off: both column capacity and placement quality significantly impact performance, and the best outcomes occur when the hardware layout and compiler jointly minimize the bridge distance. These trends align with intuition, where for fixed per-module logical work, total bridge traffic grows with the number of inter-module joint measurements, while bridge overhead scales with the average inter-column distance between paired modules. Spectral seriation reduces this average distance by embedding the interaction graph into a 1D column ordering, and larger column capacity reduces the fraction of pairings that must traverse multiple columns, lowering distance and contention.

\begin{figure}[t]
    \centering
    \includegraphics[width=0.98\columnwidth]{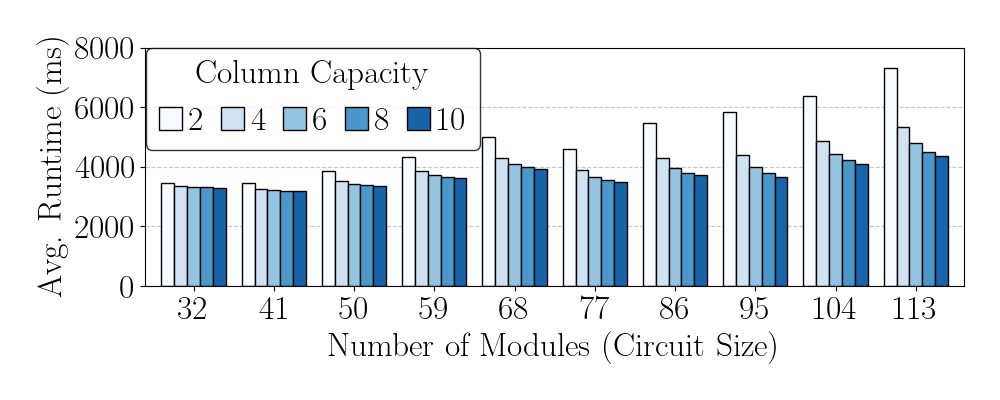}
    \vspace{-5mm}
    \caption{Average estimated runtime using spectral placement, grouped by number of modules and then column capacity.}
    \label{fig:spectral_by_size}
    \vspace{-2mm}
\end{figure}

\paragraph{Column Capacity Impact} Fig.~\ref{fig:spectral_by_size} shows our results using only the spectral placer while varying the compute-column capacity. For all column capacities, the estimated runtime increases with circuit size, as larger BB graphs induce more bridge traffic and longer schedules. At any fixed size, however, increasing the column capacity consistently reduces runtime: moving from a capacity of 2 to 10 yields only a modest improvement for the smallest circuits but almost a 40\% reduction for the largest (from roughly 7.3\,s down to about 4.3\,s at 113 modules). The widening gap in size indicates that the cost of short compute columns is amplified in large logical circuits, where highly connected modules are forced farther apart and generate many long-distance bridge pairs, leading to significant AOD movement interference.

\paragraph{End-to-End Runtime Sensitivity} To complement the zero-readout runtime results in Figs.~\ref{fig:by_circuit_size}-\ref{fig:spectral_by_size}, we derived end-to-end runtime from the same 150 compiled schedules per placement strategy while varying the per-cycle readout cost. As shown in Table~\ref{tab:measurement_sensitivity}, Spectral reduces end-to-end runtime by $17.9\%$ relative to arbitrary placement and $12.2\%$ relative to greedy placement at $T_{\mathrm{meas}}=0$. Even at a pessimistic $T_{\mathrm{meas}}=1500~\mu\mathrm{s}$, these reductions remain $15.6\%$ and $10.6\%$, respectively. The absolute mean savings are invariant across the sweep at $884.3$~ms versus arbitrary and $565.8$~ms versus greedy, and Spectral remains faster than arbitrary in all 150 paired schedules at $1500~\mu\mathrm{s}$.

\begin{table}[t]
    \centering
    \caption{End-to-end mean runtime under per-cycle readout costs. The last two cols. report Spectral's relative reduction.}
    \label{tab:measurement_sensitivity}
    \vspace{-1mm}
    \resizebox{\columnwidth}{!}{%
    \begin{tabular}{@{}rrrrrr@{}}
        \toprule
        $T_{\mathrm{meas}}$ & Arbitrary & Greedy & Spectral
        & vs.\ Arbitrary & vs.\ Greedy \\
        ($\mu$s) & (ms) & (ms) & (ms) & (\%) & (\%) \\
        \midrule
        0    & 4950.7 & 4632.2 & 4066.4 & 17.9 & 12.2 \\
        50   & 4974.8 & 4656.2 & 4090.5 & 17.8 & 12.2 \\
        100  & 4998.8 & 4680.3 & 4114.5 & 17.7 & 12.1 \\
        1000 & 5431.7 & 5113.2 & 4547.4 & 16.3 & 11.1 \\
        1500 & 5672.2 & 5353.7 & 4787.9 & 15.6 & 10.6 \\
        \bottomrule
    \end{tabular}%
    }
    \vspace{-4mm}
\end{table}

\paragraph{Atom-Loss Exposure}
We also post-processed the compiled schedules to measure total AOD shuttling distance $D$ and the number of SLM-AOD transfers $S$. Table~\ref{tab:atom_loss_exposure} reports the resulting expected loss, $\lambda\approx p_dD+p_sS$, at the representative setting $p_d=10^{-7}~\mu\mathrm{m}^{-1}$ and $p_s=10^{-5}$ per transfer. Spectral placement reduces the mean expected loss events from $21.9$ under arbitrary placement to $13.0$, a $40.7\%$ reduction, while greedy placement produces $18.0$ expected events. Across the full $4\times4$ parameter sweep, the mean ordering Spectral $<$ Greedy $<$ Arbitrary holds at all 16 combinations of $p_d$ and $p_s$. At the representative setting, Spectral also has no greater predicted loss than arbitrary placement in all 150 schedules.

\begin{table}[t]
    \centering
    \caption{Mean transport-induced atom-loss exposure per compiled schedule at $p_d=10^{-7}~\mu\mathrm{m}^{-1}$ and $p_s=10^{-5}$.}
    \label{tab:atom_loss_exposure}
    \vspace{-1mm}
    \begin{tabular}{@{}lrrr@{}}
        \toprule
        Placement & $D$ (m) & Transfers $S$ & Expected events $\lambda$ \\
        \midrule
        Arbitrary & 129.232 & $901{,}296$ & 21.9 \\
        Greedy    &  93.377 & $869{,}597$ & 18.0 \\
        Spectral  &  48.972 & $810{,}432$ & 13.0 \\
        \bottomrule
    \end{tabular}
    \vspace{-4mm}
\end{table}

\begin{figure}[t]
    \centering
    \includegraphics[width=0.98\columnwidth]{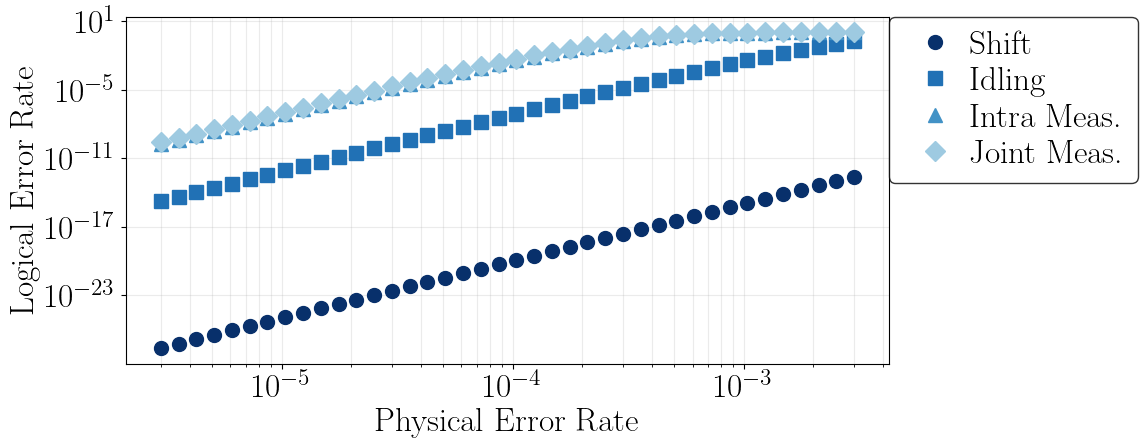}
    \vspace{-3mm}
    \caption{Logical error per BB operation vs. physical error rate.}
    \vspace{-4mm}
    \label{fig:logphyserr}
\end{figure}

\paragraph{Operational Error Impact} Fig.~\ref{fig:logphyserr} shows the logical error probability per operation as a function of the effective physical error rate. To generate it, we took a \sol{}-compiled 32-module, 6-compute-column schedule for the BB code and broke down each logical operation type (shift automorphisms, idle rounds, in-module measurements, and inter-module measurements) into four physical error groups: total qubit-shuttling distance, number of device switches, number of physical two-qubit gates, and number of measurements. We then assign physical error rates to each physical action and, for each BB operation, convert its total into an equivalent number of faults \(N_{\mathrm{ref}}\). Finally, we feed \(N_{\mathrm{ref}}\) into an oracle decoder model that declares logical failure whenever at least \(w_0 = 5\) faults occur, giving
\(
P(q) = \Pr[\mathrm{Binom}(N_{\mathrm{ref}}, q) \ge w_0]
\).

The resulting plot shows which BB-native operations are computationally expensive on neutral-atom platforms. Shift automorphisms lie many orders of magnitude below the other curves because they use no two-qubit gates and no readout, and incur only modest movement and switching overhead. Idling is more costly because each idle error-checking round already includes a nontrivial pattern of two-qubit pulses, transport, and a readout cycle. Intra-module and joint measurements lie highest on the plot since both are modeled as roughly ten idle rounds' worth of activity, and inter-module measurement adds additional bridge motion and entangling gates between bridges. Together, the curves quantify the intuition that, on neutral-atom hardware, BB-shift automorphisms are cheap while repeated syndrome-measurement cycles dominate error.

\begin{table}[t]
    \centering
    \caption{Arbitrary, Greedy, and Spectral module placement algorithms' compilation times on local hardware (in seconds).}
    \label{tab:exec_stats}
    \vspace{-1mm}
    \resizebox{\columnwidth}{!}{
        \begin{tabular}{lcccccc}
            \toprule
            & \multicolumn{3}{c}{\textbf{Circuit Size 32}} & \multicolumn{3}{c}{\textbf{Circuit Size 113}} \\
            \cmidrule(lr){2-4} \cmidrule(lr){5-7}
            \textbf{Algorithm} & \textbf{Cap=2} & \textbf{Cap=6} & \textbf{Cap=10} & \textbf{Cap=2} & \textbf{Cap=6} & \textbf{Cap=10} \\
            \midrule
            Arbitrary & 9.37 & 9.10 & 9.09 & 38.26 & 36.53 & 36.46 \\
            Greedy    & 9.39 & 9.16 & 9.11 & 39.04 & 37.95 & 37.55 \\
            Spectral  & 7.42 & 7.24 & 7.17 & 29.76 & 29.10 & 28.82 \\
            \bottomrule
        \end{tabular}
    }
    \vspace{-3mm}
\end{table}

\paragraph{Compilation Times} In Table~\ref{tab:exec_stats} we see the average time it took to run \sol{} itself on various circuits. We observe that not only does the spectral placement algorithm outperform the other techniques in the compiled runtime of the simulated program, but also in the actual runtime of \sol{} itself. Moreover, the runtime gap between the spectral placer and other techniques increases as circuit sizes grow, indicating that the spectral placer scales better. Interestingly, the spectral module placer has a lower runtime than the arbitrary placer. While the arbitrary placer doesn't incur the overhead of computing a good ordering of modules to compute columns, this leads to runtime lag because \sol{} has to compute more layers and more sequential bridge moves.

\begin{figure}[t]
    \centering
    \includegraphics[width=0.99\columnwidth]{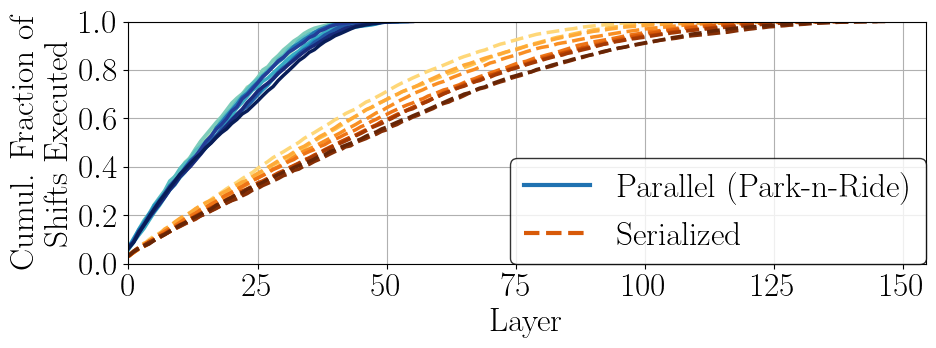}
    \vspace{-7mm}
    \caption{A comparison of how much faster \sol{} executes shift automorphisms than a technique without in-column shift parallelization. Darker colors denote larger circuits.}
    \vspace{-5mm}
    \label{fig:shift_cdf}
\end{figure}

\paragraph{Shift Automorphism Parallelization} To evaluate the impact of \sol{}'s shift automorphism parallelization, we analyze the execution rate of shift automorphisms if they were instead executed serially. Fig.~\ref{fig:shift_cdf} plots the cumulative fraction of executed shifts against the circuit layer depth, with the further left the curve sits, implying faster execution of a circuit's shift automorphisms. The results demonstrate a performance advantage for \sol{}, where its parallel approach results in shifts executing significantly faster than when executed sequentially within each compute column.

\begin{figure}[t]
    \centering
    \includegraphics[width=0.99\columnwidth]{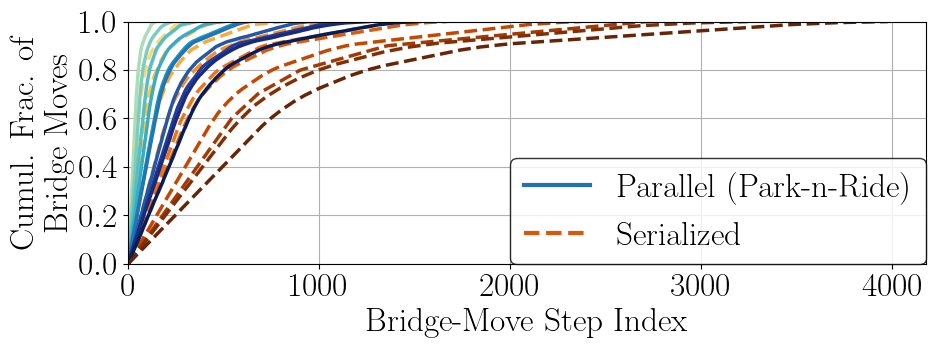}
    \vspace{-7mm}
    \caption{A comparison of the effects of parallel and sequential bridge movements to interact and execute joint measurements. Darker colors denote larger circuits.}
    \label{fig:bridge_cdf}
    \vspace{-3mm}
\end{figure}

\paragraph{Bridge Movement Parallelization} Fig.~\ref{fig:bridge_cdf} shows cumulative progress curves for bridge movement when executing joint measurements. The x-axis shows the number of micro-steps, which are move-steps for each bridge towards the interacting subzone where it will perform the bell pair that enables joint measurement. The y-axis shows the cumulative proportion of completed bridge moves executed by this circuit. Solid lines represent our parallel bridge scheduler, while dashed lines illustrate a serialized baseline that conceptually unrolls each step, allowing bridges to move one at a time. While the difference is less pronounced than with shift automorphism, it is clear that the worst-performing curves (on the right) all belong to the serialized baseline. This is because there is less parallelism to exploit with AOD constraints here since bridges often conflict with one another when moving in parallel. This is especially true for large circuits, where the number of compute columns is substantial, increasing the likelihood of AOD collisions and necessitating serialization. This further explains why the difference between small circuits of both techniques (light colors) and large circuits of both techniques (dark colors) has a much larger gap between their performance than in Fig.~\ref{fig:shift_cdf}, where column size has no effect on parallelism since shifts do not interfere with each other.

\begin{table}[t]
  \centering
  \caption{Communication scaling for inter-module joint measurements at fixed compute-column capacity (8 modules/column). We report (1) average endpoint travel-to-rendezvous distance per phase-1 joint-measure event (summed over both endpoints), (2) mean number of serialized bridge rounds required per logical layer, and (3) total bridge translation micro-steps executed by the schedule (motion-time proxy). Each entry is averaged over $n=3$ compiled instances.}
  \label{tab:bridge_scaling}
  \vspace{-1mm}
  \resizebox{0.9\columnwidth}{!}{
  \setlength{\tabcolsep}{3pt}
  \begin{tabular}{c l c c c}
    \toprule
    \makecell{\textbf{\#}\\\textbf{Modules}} &
    \makecell{\textbf{Placer}} &
    \makecell{\textbf{Avg. partner}\\\textbf{travel}} &
    \makecell{\textbf{Bridge rounds}\\\textbf{per layer}} &
    \makecell{\textbf{Bridge move}\\\textbf{micro-steps}} \\
    \midrule
    32  & Greedy   & 239.93 & 2.08  & 146.0  \\
    32  & Spectral & 197.32 & 1.57  & 100.0  \\
    68  & Greedy   & 355.82 & 5.78  & 528.0  \\
    68  & Spectral & 277.69 & 3.80  & 361.3  \\
    113 & Greedy   & 550.59 & 14.62 & 1491.3 \\
    113 & Spectral & 361.91 & 8.02  & 810.0  \\
    \bottomrule
  \end{tabular}}
  \vspace{-4mm}
\end{table}

\paragraph{Communication Results} To make communication scaling explicit, we add Table~\ref{tab:bridge_scaling}, which fixes the compute-column capacity at 8 and compares spectral versus greedy placement across representative module counts. We report three communication-facing quantities derived from the compiled motion plans: (i) an average partner distance for phase-1 inter-module measurements, which represents the mean distance each side must travel to reach the bridge rendezvous, (ii) the number of serialized bridge rounds required per logical layer (a parallelism proxy), and (iii) the total number of bridge movement micro-steps executed. Spectral placement consistently reduces distance and total bridge motion, and increases parallelism. At 113 modules, it reduces required bridge rounds per layer by 45\% and bridge micro-steps by 46\%, indicating that placement mitigates communication bottlenecks.

\paragraph{Scaling \sol{} Beyond the Gross Code}
\sol{} successfully scales beyond the $\llbracket 144,12,12\rrbracket$ code; all of its operations and optimizations remain relevant to larger BB codes. To evaluate scalability, we re-analyzed all \(450\) compiled schedules for the two-gross $\llbracket288,12,18\rrbracket$ with doubled module height and increased shift-wrap and measurement-zone travel distances. Average runtime increased by less than \(0.1\%\), while spectral placement retained its advantage (\(14.4\%\) reduction vs. arbitrary and \(10.1\%\) vs. greedy), remaining the best strategy in every run. Thus, we expect \sol{} to extend naturally to larger BB-code constructions without architectural changes.

\begin{figure}[t]
  \centering
  \includegraphics[width=0.98\columnwidth]{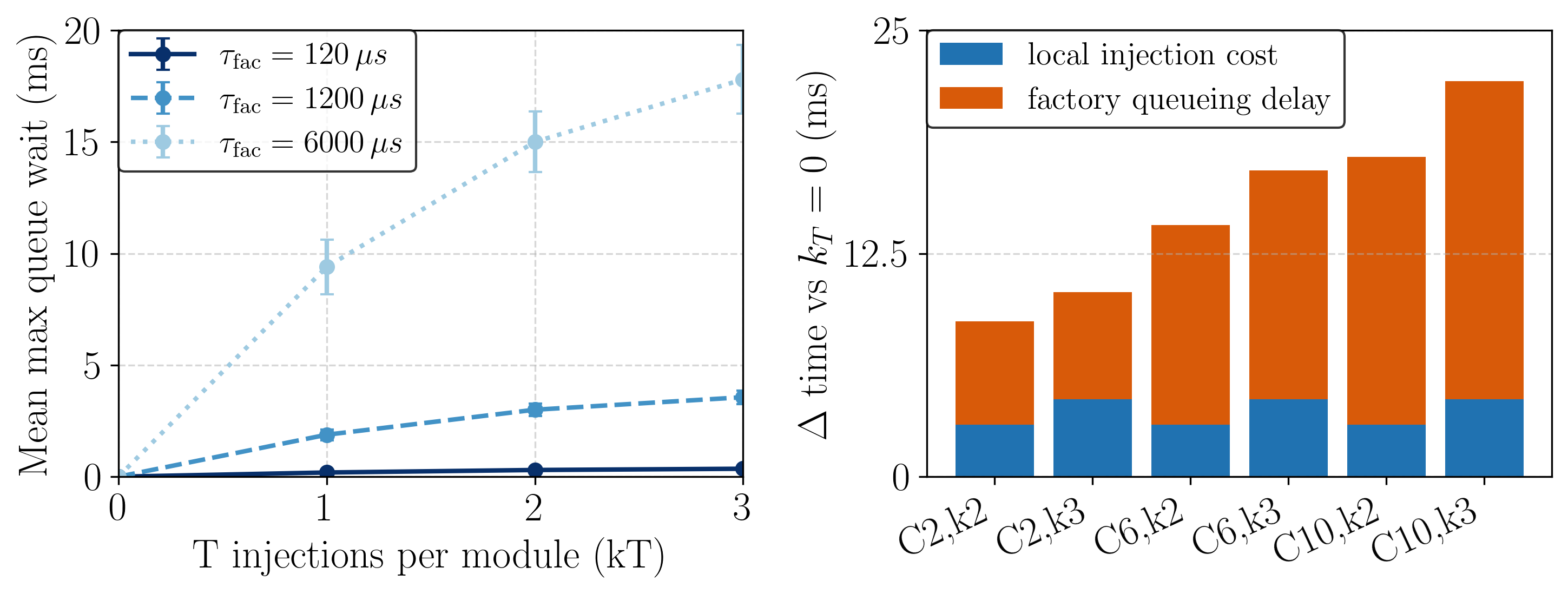}
  \vspace{-2mm}
  \caption{We insert $k_T$ synthetic T-injection requests per module and model each compute column as having a factory with time $\tau_{\mathrm{fac}}$ and local non-overlappable injection overhead $\tau_{\mathrm{inj}}$. 
  (a) Worst-case factory delay (mean max per-injection queue wait) versus $k_T$ for Cap $=10$ across three factory regimes.
  (b) At $\tau_{\mathrm{fac}}{=}6000\,\mu$s, local injection cost and worst-case factory delay for $k_T\in\{2,3\}$ across Cap $\in\{2,6,10\}$.}
  \label{fig:nonclifford_factory_pressure}
  \vspace{-5mm}
\end{figure}

\paragraph{Non-Clifford Factory Pressure} Fig.~\ref{fig:nonclifford_factory_pressure} quantifies how per-column factory throughput affects non-Clifford pressure under a per-column factory lane model. (a) plots the worst-case factory delay versus injected T demand $k_T$ for Cap $=10$ across three factory regimes. In the fast regime ($\tau_{\mathrm{fac}}{=}120\,\mu$s), contention is negligible; in the moderate regime ($\tau_{\mathrm{fac}}{=}1200\,\mu$s), delays grow to a few milliseconds by $k_T{=}3$; and in the slow regime ($\tau_{\mathrm{fac}}{=}6000\,\mu$s), worst-case delays rise to $\mathcal{O}(10)$\,ms and increase steeply with $k_T$. (b) isolates $\tau_{\mathrm{fac}}{=}6000\,\mu$s and decomposes the unavoidable local injection work ($N_T\cdot\tau_{\mathrm{inj}}$) and the worst-case factory delay. The injection work scales with $k_T$ and circuit size, while the worst-case delay increases with Cap because higher packing places more modules behind each column’s single factory server. This shows that \sol{}’s columnar organization provides a clean integration point for non-Clifford resources, and that factory provisioning (effective $\tau_{\mathrm{fac}}$ and/or factories per column) can become the dominant constraint once throughput slows.

%% file: sections/conclusion.tex
\section{Conclusion}
\label{sec:conclusion}

\sol{} enables hardware-aware compilation and execution of BB codes on zoned neutral-atom systems. By co-designing BB-native operations with movement, zoning, and interaction constraints, \sol{} translates qLDPC abstractions into valid execution schedules that maximize parallelism.  Our approach introduces parallel shift-automorphism scheduling under no-AOD-crossing constraints, spectral placement to reduce communication distance, and a bridge-based scheduler for conflict-free joint measurements. Together, these techniques reduce movement overhead and execution latency, demonstrating that BB-code primitives can be realized efficiently on reconfigurable neutral-atom architectures.

\section*{Acknowledgement}

We thank the anonymous reviewers for their helpful comments, which helped improve this work. This work was supported by the U.S. Department of Energy, Office of Science, National Quantum Information Science Research Centers, Quantum Science Center. This work was performed at Rice University and Northeastern University. Additional support was provided by the Rice Quantum Initiative, which is part of the Smalley-Curl Institute and the Ken Kennedy Institute.